\documentclass{ws-ijmpa}
\usepackage[super,compress]{cite}
\usepackage{graphicx}
\begin{document}
	\markboth{W. O. dos Santos \& E. R. B. de Mello}{Induced current in high-dimensional AdS bulk in the presence of a cosmic string and brane}
	
	\catchline{}{}{}{}{}
	%
	%%%%%%%%%%%%%%%%%%%%%%%%%%%%%%%%%%%%%%%%%%%%%%%%%%%%%%%%%%%%%%%%%%%%
	
	\title{Induced current in high-dimensional AdS bulk in the presence of a cosmic string and brane}
	
	\author{W. Oliveira dos Santos and E.  R. Bezerra de Mello}
	
	\address{Dept. de Física-CCEN, Universidade Federal da Paraíba, 58.059-970\\
		João Pessoa, Paraíba, Caixa Postal 5.008, Brazil}

	\maketitle
	
	\begin{history}
		\received{Day Month Year}
		\revised{Day Month Year}
	\end{history}
	
	\begin{abstract}
In this paper we analyze the vacuum bosonic current associated with a massive charged scalar field propagating 	in a $(D+1)$-dimensional anti-de Sitter (AdS) bulk in the presence of an idealized carrying-magnetic-flux cosmic string and also a brane. We assume that the string is perpendicular to the brane which is parallel to the AdS boundary. 
 The brane divides the space in two regions with different	properties of vacuum states. We investigate the only non-vanishing azimuthal components of the induced currents in both regions. To develop this analysis we calculate first, for both regions, the positive frequency Wightman functions. These present  contributions associated with the presence of a cosmic string only plus the other induced by the brane. In this talk we consider only the contributions induced by the brane.
		
	\end{abstract}
	
	\ccode{PACS numbers: 98.80.Cq, 11.10.Gh, 11.27.+d}
	
	%\tableofcontents
	\section{Introduction}
	\label{Int}
	Cosmic strings are linear topological objects that may be formed as consequence of gauge symmetry breaking in early stage of the Universe evolution \cite{Kibble,V-S}. They produce planar angle deficits in the two-dimensional sub-space orthogonal to them. 
	
	The anti-de-Sitter (AdS) space is a maximally symmetric solution of the Einstein equation in presence of negative cosmological constant. The importance of  this model increased when it was discovered that AdS spacetime arises as a ground state in extended supergravity
	and in string theories.
	
	The investigation of the geometry of the spacetime considering the presence of a cosmic string in AdS bulk was developed in  Refs. \refcite{Ghe1,Cristine}. There it was observed that at distances bigger than the string's core radius, the effect produced by the string can be  described by a planar deficit angle in the AdS metric, similarly to the case in the Minkowski spacetime. The combined effect of curvature and conical topology produce non-vanishing results in the evaluations of vacuum expectation values of several physical observable, as the energy-momentum tensor and current densities.  
	
	Here we want to investigate the vacuum bosonic current induced by a magnetic flux running along the cosmic string in a $(D+1)$-dimensional AdS bulk considering the presence of a flat brane parallel to the AdS boundary. We assume that on the brane the charged bosonic field obeys the Robin boundary condition. The brane separates the space in two regions with different	properties of vacuum states. We want to investigate the azimuthal component of the induced current in both regions. This is will done by calculating the corresponding positive frequency Wightman functions. These  present contributions associated with the  cosmic string without brane, plus the other induced by the brane. Ou main objective is to consider only the contributions induced by the brane.

	\section{The Klein-Gordon equation}
	\label{K-G}
	In cylindrical coordinates, and using the {\it Poincaré} coordinate the geometry associated with a cosmic string in $(1+D)$-AdS spacetime is given by the line element below
	\begin{equation}
		\label{ds2_gen}
		ds^2 = \left(\frac{a}{w}\right)^2[dt^2 - dr^2 - r^2d\varphi^2 - dw^2 - \sum_{i=4}^{D}(dx^i)^2]   \   ,
	\end{equation}
where  $r\geq 0$, $\phi \in \lbrack 0,\ 2\pi /q]$\footnote{ The parameter  $q\geq 1 $ encodes the conical topology of the spacetime} and $x^i\in (-\infty ,\ \infty )$. The variable $w$, is defined in the interval $\lbrack 0,\ \infty )$. Two specific values for it are: $w=0$ (AdS boundary) and $w=\infty$ (horizon). 
	
	The field equation that governs the relativistic quantum motion of a charged bosonic field with mass $m$, in a curved background and in the presence of an electromagnetic potential vector, $A_\mu$, is give by,
	\begin{equation}		
		(g^{\mu\nu}D_{\mu}D_{\nu} + m^2 + \xi R)\varphi(x) = 0  \   , 
		\label{KGE}
	\end{equation}
	with $D_{\mu}=\nabla_{\mu}+ieA_{\mu}$. In the above equation we have introduced a non-minimal coupling $\xi R$, between the field and the geometry.
	
	In this paper we will consider the case of an idealized carrying-magnetic-flux cosmic string, i.e., a very thin cosmic string presenting a magnetic flux running along its core. The corresponding magnetic flux is represented by the following vector potential
	\begin{eqnarray}
		\label{Vector_Pot}
		A_{\mu} =-\frac{q\Phi}{2\pi}\delta_\mu^\phi  \ , 
	\end{eqnarray}
	where $\Phi$ represents the total magnetic flux. We also admit the presence of a flat brane located at $w=w_0$ . On the brane the field operator obeys the Robin condition,
	\begin{eqnarray}
		\label{RBC}
		(1+\beta n^{\mu}{D}_{\mu
		})\varphi(x)=0, \ w=w_0 \ .
	\end{eqnarray}
	The vector $n^{\mu}$ is the inward pointing normal to the brane at $w=w_0$. In the region $0\le w\le w_0$,  $L$ (left)-region, this normal is given as $n^{\mu}=-\delta_{3}^{\mu}a/w$ and in the region $w_0\le w\le \infty$, $R$ (right)-region, as  $n^{\mu}=\delta_{3}^{\mu}a/w$.
	
	In the geometry under consideration and taking into account the azimuthal vector potentials, (\ref{Vector_Pot}), the Klein-Gordon equation (KG), Eq. (\ref{KGE}), is written as,
	\begin{eqnarray}
		\label{KGE2}
		&&\left[\frac{\partial^2}{\partial t^2} - \frac{\partial^2}{\partial r^2} - \frac{1}{r}\frac{\partial}{\partial r} - \frac{1}{r^2}\left(\frac{\partial}{\partial\phi} + ieA_{\phi}\right)^2
		- \frac{\partial^2}{\partial w^2}\right.\nonumber\\
		&&\left.-\frac{(1-D)}{w}\frac{\partial}{\partial w} + \frac{M(D,m,\xi)}{w^2} - \sum_{i=4}^{D}\frac{\partial^2}{\partial (x^i)^2} \right]\varphi(x) = 0  \  , 
	\end{eqnarray}
	with $M(D,m,\xi) = a^2m^2 - \xi D(D+1)$.
	
	Considering previous analysis \cite{Wagner_19}, we can express the positive energy solution as:
	\begin{equation}
		\varphi_{\sigma}(x) = C_{\sigma}w^{\frac{D}{2}}W_{\nu}(pw)J_{q|n +\alpha|}(\lambda r)e^{- iE t + iqn\phi + i\vec{k}\cdot\vec{x}_{\parallel}} \ ,
		\label{Solu1}
	\end{equation}
	being $W_\nu(pw)$ a linear combination of the Bessel and Neumann functions \cite{Grad}:
	\begin{equation}
		W_{\nu}(pw)=C_1J_{\nu}(pw)+C_2Y_{\nu}(pw) \  , \ \nu = \sqrt{{D^2}/{4}+M(D,m,\xi)} \  .
		\label{W-function}
	\end{equation}
	
	As to the energy, $E$, and the parameter $\alpha$ in the order of the Bessel function associated with the radial coordinate, we have:
	\begin{eqnarray}
		E =\sqrt{\lambda^2 + p^2 + \vec{k}^2}  \ , \ {\rm and} \ 
		\alpha = \frac{eA_{\phi}}{q} = -\frac{\Phi_{\phi}}{\Phi_0} \ , 
		\label{const}
	\end{eqnarray}
	with $\Phi_0=\frac{2\pi}{e}$. In (\ref{Solu1}) $\vec{x}_{\parallel}$ represents the coordinates defined in the $(D-3)$ extra dimensions, and $\vec{k}$ the corresponding momentum.  The indices $\sigma$ represents the set of quantum numbers $(n, \lambda, p, \vec{k})$, where $n=0,\pm1,\pm2,\ldots$, $\lambda \geq 0$, $-\infty<k^j<\infty$ for $j=4,...,D$. The quantum number $p$, is determined, separately, for each region.
	
	The normalization coefficient, $C_{\sigma}$,  is obtained by the condition
	\begin{eqnarray}
		\int d^Dx\sqrt{|g|}g^{00}\varphi_{\sigma'}^{*}(x)\varphi_{\sigma}(x)= \frac{1}{2E}\delta_{\sigma,\sigma'}  \   ,
		\label{NC}
	\end{eqnarray}
where delta symbol on the right-hand side is understood as Dirac delta function for the continuous quantum number, and Kronecker delta for the discrete one.

	\subsection{Normalized wave-functions in $R$-region}	
	Let us consider first the $R$-region. The Robin condition provides a relation between the coefficients $C_1$ and $C_2$ in (\ref{W-function}), given by  $C_2/C_1=-\bar{J}_{\nu}(pw_0)/\bar{Y}_{\nu}(pw_0)$; where  we use the compact notation
	\begin{equation}
		\label{F_function}
		\bar{F}(x)=A_{0}F(x)+B_{0}xF^{\prime}(x) \ ,
	\end{equation}
	with the coefficients $	A_{0}=1+\frac{D\beta}{2a}$ and $B_{0}=\beta/a$.
	
	Using this notation, the normalized solutions of KG equation that obeys the boundary condition can be expressed by,
	\begin{equation}
		\varphi_{(R)\sigma}(x) = C_{(R)\sigma}w^{\frac{D}{2}}g_{\nu}(pw_0,pw)J_{q|n +\alpha|}(\lambda r)e^{- iE t + iqn\phi + i\vec{k}\cdot\vec{x}_{\parallel}} \ .
		\label{Solu-R}
	\end{equation}
	In the above equation we have introduced the function
	\begin{equation}
		\label{g-function}
		g_{\nu}(u,v)=J_{\nu}(v)\bar{Y}_{\nu}(u)-\bar{J}_{\nu}(u)Y_{\nu}(v) \ .
	\end{equation}
	In this region the quantum number $p$ assumes  continuous values; so the normalization constant is
	\begin{equation}
		|C_{(R)\sigma}|^2=\frac{(2\pi)^{2-D}qp\lambda}{2Ea^{D-1}[\bar{J}_{\nu}^2(pw_0)+\bar{Y}_{\nu}^2(pw_0)]} \ .
		\label{C_R-region}
	\end{equation}
	
	\subsection{Normalized wave-functions in $L$-region}
	
	In the $L$-region, the integration over $w$ in (\ref{NC}), is restricted in the interval $0\le w\le w_0$. Assuming the Dirichlet boundary condition on $w=0$ we have $C_2=0$. Thus, with this condition, the mode functions are given by
	\begin{equation}
		\varphi_{(L)\sigma}(x) = C_{(L)\sigma}w^{\frac{D}{2}}J_{\nu}(pw)J_{q|n +\alpha|}(\lambda r)e^{- iE t + iqn\phi + i\vec{k}\cdot\vec{x}_{\parallel}} \ .
		\label{Solu-L}
	\end{equation}
	According to the Robin boundary condition (\ref{RBC}), the eigenvalues of the quantum number $p$ must obey a restriction given by:
	\begin{equation}
		\label{J_barr}
		\bar{J}_{\nu}(pw_0)=0 \ ,
	\end{equation}
	where $	\bar{J}_{\nu}(x)$  was given by (\ref{F_function}), with $	A_{0}=1-\frac{D\beta}{2a}$ and $B_{0}=-\beta/a $.
	
	Considering that eigenvalues of (\ref{J_barr}) are given by $p=p_{\nu,i}/w_0$, with $p_{\nu,i}$,  the positive zeros of the function $\bar{J}_{\nu}(x)$  are enumerated by $i=1, 2,...$. Taking now the normalization condition (\ref{NC}), with $\delta_{p,p^\prime}=\delta_{i,i^\prime}$, we obtain
	\begin{equation}
		|C_{(L)\sigma}|^2=\frac{(2\pi)^{2-D}qp_{\nu,i}\lambda T_{\nu}(p_{\nu,i})}{w_0a^{D-1}\sqrt{p_{\nu,i}^2+w_{0}^2(\lambda^2+\vec{k}^2})}\ .
		\label{C_L-region}
	\end{equation}
	Where in the above expression we have defined the function $T_\nu(z)$ by,
	\begin{eqnarray}
		T_{\nu}(z)=z[(z^2-\nu^2)J_{\nu}^2(z)+z^2(J_{\nu}^{\prime}(z))^2]^{-1} \ .
	\end{eqnarray}

	\section{Wightman Function}
	\label{Wight_fun}
	Our main objective in this section is to obtain the positive frequency Wightman functions induced by the brane for both regions, $R$ and $L$.  These functions can be obtained by sum over the  complete set of bosonic modes, as shown below: 
	\begin{equation}
		W(x,x^{\prime})=\sum_{\sigma}\varphi_{\sigma}(x)\varphi_{\sigma}^{\ast}(x^{\prime}) \ .
		\label{Wightman-def}
	\end{equation}
	\subsection{$R$-region} 
	Combining previous results, (\ref{Solu-R}), (\ref{g-function}) and (\ref{C_R-region}), we obtain  the normalized positive energy solution of the KG equation. Substituting this function into (\ref{Wightman-def}), we obtains:
	\begin{eqnarray}
		W_R(x,x^{\prime})&=&\frac{q(ww')^{D/2}}{2a^{D-1}(2\pi)^{D-2}}\sum_\sigma\frac{p\lambda}E\frac{g_{\nu}(pw_0,pw)g_{\nu}(pw_0,pw^\prime)}{(\bar{J}_{\nu}^2(pw_0)+\bar{Y}_{\nu}^2(pw_0))}J_{q|n +\alpha|}(\lambda r)J_{q|n +\alpha|}(\lambda r^\prime)\nonumber\\&\times&e^{- iE(t-t^\prime) + iqn(\phi-\phi^\prime) + i\vec{k}\cdot(\vec{x}_{\parallel}- \vec{x}^\prime_{\parallel})} \  .  \label{W_R}
	\end{eqnarray}
	We use the compact notation below for the summation over $\sigma$ in (\ref{W_R}):
	\begin{equation}
		\sum_{\sigma }=\sum_{n=-\infty}^{+\infty} \ \int dp \ \int_0^\infty
		\ d\lambda \ \int d{\vec{k}} \ .  \label{Sumsig}
	\end{equation}
	Substituting $E=\sqrt{\lambda^2+p^2+\vec{k}^2}$, and using the identity,
	\begin{equation}
		\frac{e^{-E\Delta \tau}}{E}=\frac2{\sqrt{\pi}}\int_0^\infty ds e^{-s^2E^2-(\Delta \tau)^2/(4s^2)}  \  ,
		\label{identity}
	\end{equation}
	we can integrate over $\lambda$ and $\vec{k}$  by using \cite{Grad}, resulting in
	\begin{eqnarray}
		W_R(x,x^\prime)&=&\frac{qrr^\prime}{2(2\pi)^{D/2}a^{D-1}}\left(\frac{ww^\prime}{rr^\prime}\right)^{D/2}\int_{0}^{\infty}dv v^{\frac{D}{2}-2}e^{-\frac{\rho^2}{2rr^\prime}v}\nonumber\\
		&\times&\sum_{n}e^{inq\Delta\phi}I_{q|n+\alpha|}(v)\int_{0}^{\infty}dppe^{-\frac{rr^\prime}{2v}p^2}\frac{g_{\nu}(pw_0,pw)g_{\nu}(pw_0,pw^{\prime})}{\bar{J}_{\nu}^2(pw_0)+\bar{Y}_{\nu}^2(pw_0)} \ ,
		\label{W-function_R}
	\end{eqnarray}
	where we have used the notation $\rho^2=r^2+r^{\prime2}+\Delta\vec{x}_{\parallel}^2-\Delta t^2$.
	
	\subsection{$L$-Region}
	To obtain the positive frequency Wightman function in the $L$-region, we have to substitute the expression (\ref{Solu-L}), with the coefficient (\ref{C_L-region}), into (\ref{Wightman-def}). So we have:
	\begin{eqnarray}
		W_L(x,x^\prime)&=&\frac{q(ww^\prime)^{D/2}}{a^{D-1}(2\pi)^{D-2}w_0^2}\sum_{\sigma}\frac{\lambda p_{\nu,i}}{\sqrt{(p_{\nu,i}/w_0)^2+\lambda^2+\vec{k}^2}}T_{\nu}(p_{\nu,i})J_{\nu}(p_{\nu,i}w)J_{\nu}(p_{\nu,i}w^\prime)
		\nonumber\\
		&\times&J_{q|n+\alpha|}(\lambda r)J_{q|n+\alpha|}(\lambda r^\prime)e^{inq\Delta\varphi+i\vec{k}\cdot\Delta \vec{x}-iE\Delta t} \ .
	\end{eqnarray}
	However, for this region the summation over $\sigma$ is given by,
	\begin{equation}
		\sum_{\sigma}=\sum_{n=-\infty}^{+\infty}\int_{0}^{\infty}d\lambda\sum_{i=1}^{\infty}\int d\vec{k} \ .
	\end{equation}
	
	Using again the identity (\ref{identity}),  we can integrate over $\lambda$ and $\vec{k}$ with the help of \cite{Grad}, and obtain the expression:
	\begin{eqnarray}
		\label{W_L}
		W_L(x,x^\prime)&=&\frac{q(ww')^{D/2}a^{1-D}}{(2\pi)^{D/2}w_0^2(rr')^{D/2-1}}\int_{0}^{\infty}dvv^{D/2-2}
		e^{-\frac{\rho^2}{2rr^\prime}v}
		\sum_{n=-\infty}^{\infty}e^{inq\Delta\varphi}I_{q|n+\alpha_0|}(v)	\nonumber\\
		&\times&\sum_{i=1}^{\infty}p_{\nu,i}T_{\nu}(p_{\nu,i})J_{\nu}(p_{\nu,i}w/w_0)J_{\nu}(p_{\nu,i}w'/w_0)e^{-\frac{r^2p_{\nu,i}^2}{2w_0^2v}} \ .
	\end{eqnarray}

	\subsection{Boundary induced Wightmann function}
	The boundary induced Wightman function in $R$ and $L$ regions, $	W_{b(R,L)}(x,x^\prime)$, can be obtained by subtraction form the previous results, Eqs. (\ref{W-function_R}) and (\ref{W_L}), the function associated with the cosmic string in AdS without boundary, as shown below:
	{\begin{eqnarray}
			W_{b(R,L)}(x,x^\prime)=W_{(R,L)}(x,x^\prime)-W_{cs}(x,x^\prime) \ .
		\end{eqnarray}
		\subsection{R-region}
		The Wightman function in $R$-region can be obtained by using the identity:
		\begin{eqnarray}
			\frac{g_{\nu}(u,v)g_{\nu}(u,v^{\prime})}{\bar{J}_{\nu}^2(u)+\bar{Y}_{\nu}^2(u)}-
			J_{\nu}(v)J_{\nu}(v^\prime)=-\frac{1}{2}\sum_{l=1}^{2}\frac{\bar{J}_{\nu}(u)}{\bar{H}_{\nu}^{(l)}(u)}H_{\nu}^{(l)}(v)H_{\nu}^{(l)}(v^\prime) \ ,
		\end{eqnarray}
		where $H_{\nu}^{(l)}(x)$, $l=1, 2$, represent the Hankel functions \cite{Grad}. After some intermediate steps, we get
		\begin{eqnarray}
			W_{b(R)}(x,x^\prime)&=&-\frac{qa^{1-D}rr^\prime}{4(2\pi)^{D/2}}\left(\frac{ww^\prime}{rr^\prime}\right)^{D/2}\int_{0}^{\infty}dv v^{\frac{D}{2}-2}e^{-\frac{\rho^2}{2rr^\prime}v}\sum_{n}e^{inq\Delta\phi}I_{q|n+\alpha|}(v)\nonumber\\
			&\times&\int_{0}^{\infty}dppe^{-\frac{rr^\prime}{2v}p^2}\sum_{l=1}^{2}\frac{\bar{J}_{\nu}(pw_0)}{\bar{H}_{\nu}^{(l)}(pw_0)}H_{\nu}^{(l)}(pw)H_{\nu}^{(l)}(pw^\prime) \ .
			\label{W-function_R_b}
		\end{eqnarray}
		The parameter $\alpha$ can be written in the form
		\begin{eqnarray}
			\alpha=n_{0}+\alpha_0, \ \textrm{with}\ |\alpha_0|<\frac{1}{2},
			\label{const-2}
		\end{eqnarray}
		with $n_{0}$ being an integer number. The sum over the quantum number $n$ has been developed in Ref. \refcite{deMello:2014ksa}, providing the following result:
		\begin{eqnarray}
			&&\sum_{n=-\infty}^{\infty}e^{iqn\Delta\phi}I_{q|n+\alpha|}(v)=\frac{1}{q}\sum_{k}e^{v\cos(2\pi k/q-\Delta\phi)}e^{i\alpha(2\pi k -q\Delta\phi)}-\frac{e^{-iqn_{0}\Delta\phi}}{2\pi i}\\
			&&\times\sum_{j=\pm1}je^{ji\pi q|\alpha_0|}
			\int_{0}^{\infty}dy\frac{(\cosh{[qy(1-|\alpha_0|)]}-\cosh{(|\alpha_0| qy)e^{-iq(\Delta\phi+j\pi)}})}{e^{v\cosh(y)}(\cosh{(qy)}-\cos{(q(\Delta\phi+j\pi))}} \ .
		\end{eqnarray}
		With the parameter $k$ varying  in the interval:
		\begin{eqnarray}
			-\frac{q}{2}+\frac{\Delta\phi}{\Phi_{0}}\le k\le \frac{q}{2}+\frac{\Delta\phi}{\Phi_{0}}  \   .
		\end{eqnarray}
		Finally we rotate the contour integration over $p$ by the angle $\pi/2$ $(-\pi/2)$ for the term $l=1$ $(l=2)$. Using the relations involving Bessel functions, the result is
		\begin{eqnarray}
			&&W_{b(R)}(x,x^\prime)=-\frac{(ww^\prime)^{D/2}}{(2\pi)^{D/2}a^{D-1}}\int_{0}^{\infty}dpp^{D-1}\frac{\bar{I}_{\nu}(pw_0)}{\bar{K}_{\nu}(pw_0)}K_{\nu}(pw)K_{\nu}(pw^\prime)\nonumber\\
			&&\times\Biggl\{\sum_{k}e^{i\alpha(2\pi k -q\Delta\phi)}f_{\frac{D}{2}-1}(pu_k)-\frac{qe^{-iqn_{0}\Delta\phi}}{2\pi i}\sum_{j=\pm1}je^{ji\pi q|\alpha_0|}
			\nonumber\\
			&&\times\int_{0}^{\infty}dy\frac{\cosh{[qy(1-|\alpha_0|)]}-\cosh{(|\alpha_0| qy)e^{-iq(\Delta\phi+j\pi)}}}{\cosh{(qy)}-\cos{(q(\Delta\phi+j\pi))}}f_{\frac{D}{2}-1}(pu_y)\Biggl\} \ ,
			\label{W-function_b-2}
		\end{eqnarray}
where we have defined $	f_{\mu}(x)={J_{\mu}(x)}/{x^{\mu}} $ and introduced the notations
		\begin{eqnarray}
			\label{u_function}
			u_{k}^2&=&r^2+r'^2-2rr'\cos{(2\pi k/q-\Delta\phi)}
			+\Delta\vec{x}_{\parallel}^{2}-\Delta t^2\ , \nonumber\\
			u_{y}^2&=&r^2+r'^2+2rr'\cosh{(y)}+\Delta\vec{x}_{\parallel}^{2}-\Delta t^2 \ .
		\end{eqnarray}

The 		
		\subsection{L-region} 
		The Wightman function in $L-$region is obtained in different way. The reason is because the quantum number associated with the momentum along the {\it Poincar\'e} coordinate is discrete. So the summation over this number can be evaluated by using the generalized Abel-Plana summation formula
		\cite{SahaRev}
		\begin{equation}
			\sum_{i=1}^{\infty}T_{\nu}(p_{\nu,i})f(p_{\nu,i})=\frac{1}{2}\int_{0}^{\infty}dzf(z)-\frac{1}{2\pi}\int_{0}^{\infty}dz\frac{\bar{K}_{\nu}(z)}{\bar{I}_{\nu}(z)}\left[e^{-i\nu z}f(iz)+e^{i\nu z}f(-iz)\right] \ .
			\label{Abel-Plana}
		\end{equation}
		For the problem that we are analyzing the function $f(z)$ is given below,
		\begin{equation}
			f(z)=z^{D/2}J_{\nu}(zw/w_0)J_{\nu}(zw^\prime/w_0)K_{\frac{D}{2}-1}(2uz/w_0) \ .
			\label{func}
		\end{equation}
		
		The first term given by (\ref{Abel-Plana}) corresponds the Wightman function in the absence of brane. So our interest here resides in the second term. From now on we adopt a procedure analogous to that was developed in the previous calculation.\cite{DosSantos:2023ord} The final result is:
		\begin{eqnarray}
			&&W_{b(L)}(x,x^\prime)=-\frac{(ww^\prime)^{D/2}}{(2\pi)^{D/2}a^{D-1}}\int_{0}^{\infty}dpp^{D/2}\frac{\bar{K}_{\nu}(pw_0)}{\bar{I}_{\nu}(pw_0)}I_{\nu}(pw)I_{\nu}(pw^\prime)\nonumber\\
			&&\times\Biggl\{\sum_{k}e^{i\alpha(2\pi k -q\Delta\phi)}f_{\frac{D}{2}-1}(pu_k)-\frac{qe^{-iqn_{0}\Delta\phi}}{2\pi i}\sum_{j=\pm1}je^{ji\pi q|\alpha_0|}
			\nonumber\\
			&&\times\int_{0}^{\infty}dy\frac{\cosh{[qy(1-|\alpha_0|)]}-\cosh{(|\alpha_0| qy)e^{-iq(\Delta\phi+j\pi)}}}{\cosh{(qy)}-\cos{(q(\Delta\phi+j\pi))}}f_{\frac{D}{2}-1}(pu_y)\Biggl\} 
			\label{W-function_b-3-L} \  .
		\end{eqnarray}
		In (\ref{W-function_b-3-L}) we adopt the notation given in (\ref{u_function}).
		
		\section{Boundary induced azimuthal current}
		The boundary induced bosonic current can be obtained by using the positive frequency Wightman function. So, for the system under consideration, it reads:
		\begin{eqnarray}
			\label{current}
			\langle j_\phi(x)\rangle_{b(R.L)}=ie\lim_{x'\to x}[(\partial_\phi-\partial_{\phi'})W_{b(R,L)}(x',x)+2ieA_\phi W_{b(R,L)}(x',x)] \  .
		\end{eqnarray}

		\subsection{R-region}
		In the $R$-region the current can be obtained by substituting (\ref{W-function_b-2}) into (\ref{current}). After proceeding with some intermediate calculations, we obtain:
		\begin{eqnarray}
			\label{azimu_brane_R}
			\langle j^{\phi}\rangle_{b(R)}&=&-\frac{4e}{(2\pi)^{D/2}a^{D+1}}\int_{0}^{\infty}dzz^{D+1}\frac{\bar{I}_{\nu}(zw_0/w)}{\bar{K}_{\nu}(zw_0/w)}K_{\nu}^2(z)\nonumber\\
			&\times&\Bigg[{\sum_{k=1}^{[q/2]}}\sin{(2\pi k/q)}\sin{(2\pi k\alpha_0)}f_{\frac{D}{2}}(2z(r/w)s_k)
			\nonumber \\
			&+&\frac{q}{2\pi}\int_{0}^{\infty}d\eta\frac{\sinh{(\eta)}g(q,\alpha_0,\eta)}{\cosh(q\eta)-\cos(q\pi)}f_{\frac{D}{2}}(2z(r/w)\cosh(\eta/2))\Bigg] \ ,
		\end{eqnarray}
where $[q/2]$ represents the integer part  of $q/2$. For $q<2$ the summation contribution is absent.

Having obtained the above result, our main interest is to investigate its behavior for points very far from the brane, i.e., for $w/w_0\gg1$. For this case, we obtain a power-type decay as exhibited below:
\begin{eqnarray}
\label{azimu_brane_R_1}
&&	\langle j^{\phi}\rangle_{b(R)}\approx- \frac{2^{3-2\nu-D/2}e}{\pi^{D/2}\Gamma(\nu)\Gamma(\nu+1)a^{D+1}}\left(\frac{A_0+\nu B_0}{A_0-\nu B_0}\right)\left(\frac{w_0}{w}\right)^{2\nu} \nonumber\\
&\times&\int_{0}^{\infty}dzz^{D+2\nu+1}K_{\nu}^2(z)\Bigg[\sum_{k=1}^{[q/2]}\sin{(2\pi k/q)}\sin{(2\pi k\alpha_0)}f_{\frac{D}{2}}(2z(r/w)s_k)
\nonumber \\
&+&\frac{q}{2\pi}\int_{0}^{\infty}dy\frac{\sinh{(y)}g(q,\alpha_0,y)}{\cosh(qy)-\cos(q\pi)}f_{\frac{D}{2}}(2z(r/w)\cosh(y/2))\Bigg] \ .
\end{eqnarray}

\subsection{L-region}
	The boundary induced current in the $L$-region, can be obtained following similar procedure as before. Omitting these intermediate steps we get, 
	\begin{eqnarray}
		\label{azimu_brane_L}
		\langle j^{\phi}\rangle_{b(L)}&=&-\frac{4e}{(2\pi)^{D/2}a^{D+1}}\int_{0}^{\infty}dzz^{D+1}\frac{\bar{K}_{\nu}(zw_0/w)}{\bar{I}_{\nu}(zw_0/w)}I_{\nu}^2(z)\nonumber\\
		&\times&\Bigg[\sum_{k=1}^{[q/2]}\sin{(2\pi k/q)}\sin{(2\pi k\alpha_0)}f_{\frac{D}{2}}(2z(r/w)s_k)
		\nonumber \\
		&+&\frac{q}{2\pi}\int_{0}^{\infty}d\eta\frac{\sinh{(\eta)}g(q,\alpha_0,\eta)}{\cosh(q\eta)-\cos(q\pi)}f_{\frac{D}{2}}(2z(r/w)\cosh(\eta/2))\Bigg] \ .
\end{eqnarray}

Proceeding similarly as in the $R$-region, we now investigate the current for points near the AdS boundary ($w = 0$), $w\ll w_0$. So, we get,
	\begin{eqnarray}
		\label{azimu_brane_L-asymp}
		\langle j^{\phi}\rangle_{b(L)}&\approx&-\frac{2^{2-2\nu-D/2}e}{\pi^{D/2}\Gamma^2(\nu+1)a^{D+1}}\left(\frac{w}{w_0}\right)^{2\nu}\int_{0}^{\infty}dzz^{D+2\nu+1}\frac{\bar{K}_{\nu}(zw_0/w)}{\bar{I}_{\nu}(zw_0/w)}\nonumber\\
		&\times&\Bigg[\sum_{k=1}^{[q/2]}\sin{(2\pi k/q)}\sin{(2\pi k\alpha_0)}f_{\frac{D}{2}}(2z(r/w)s_k)
		\nonumber \\
		&+&\frac{q}{2\pi}\int_{0}^{\infty}d\eta\frac{\sinh{(\eta)}g(q,\alpha_0,\eta)}{\cosh(q\eta)-\cos(q\pi)}f_{\frac{D}{2}}(2z(r/w)\cosh(\eta/2))\Bigg] \ . 
	\end{eqnarray}

\section{Total induced azimuthal current}
	\label{total}
The total induced azimuthal current is given by adding the current in AdS space without brane plus the brane-induced contribution. The current induced by the cosmic string only has been calculated in \cite{Wagner_19}, and it reads,
\begin{eqnarray}
	\langle j^{\phi}(x)\rangle_{cs}&=&\frac{4ea^{-(1+D)}}{(2\pi)^{\frac{D+1}{2}}}\Bigg[\sum_{k=1}^{[q/2]}\sin{(2\pi k/q)}\sin{(2\pi k\alpha_0)}F_{\nu-1/2}^{(D+1)/2}(u_{k})\nonumber\\
	&+&\frac{q}{\pi}\int_{0}^{\infty}dy\frac{\sinh{(2y)}g(q,\alpha_0,2y)}{\cosh{(2qy)}-\cos{(\pi q)}}F_{\nu-1/2}^{(D+1)/2}(u_{y})\Bigg] \ ,
\end{eqnarray}
where the arguments of the functions are given by
\begin{eqnarray}
	u_{k}=1+2(r/w)^2\sin^2{(\pi k/q)} \ {\rm and} \	u_{y}=1+2(r/w)^2\cosh^2{(y)}  \  ,
\end{eqnarray} 
being
\begin{eqnarray}
F_{\gamma}^{\mu}(u)=\frac{\sqrt{\pi}\Gamma(\gamma+\mu+1)}{2^{\gamma+1}\Gamma(\gamma+3/2)u^{\gamma+\mu+1}}F\bigg(\frac{\gamma+\mu}{2}+1,\frac{\gamma+\mu+1}{2};\gamma+\frac{3}{2};\frac{1}{u^{2}}\bigg) \ ,
\end{eqnarray}
where $
F(a,b;c;z)$ corresponds to hypergeometric function \cite{Grad}.

In order to compare the the magnitude of the current induced by the cosmic string in the absence of brane, $\langle j^{\phi}(x)\rangle_{cs}$, with the one induced by the  brane, $\langle j^{\phi}\rangle_{b(R,L)}$, the graph on the top of  Fig.\ref{fig1} presents their behaviors as function of $r/w$ for the $R-$region, considering $D=3$, $\nu=3/2$, $\alpha_0=0.4$ and $q=1.5$. For the $L-$region, their behaviors are presented in the graph on the bottom, adopting the same values for the parameters. For both brane-induced currents, we considered, separately, Dirichlet and Neumann boundary conditions. By these plots, we can observe that near the string, $\langle j^{\phi}\rangle_{cs}$ is dominant.
\begin{figure}[!htb]
	\begin{center}
		\includegraphics[scale=0.30]{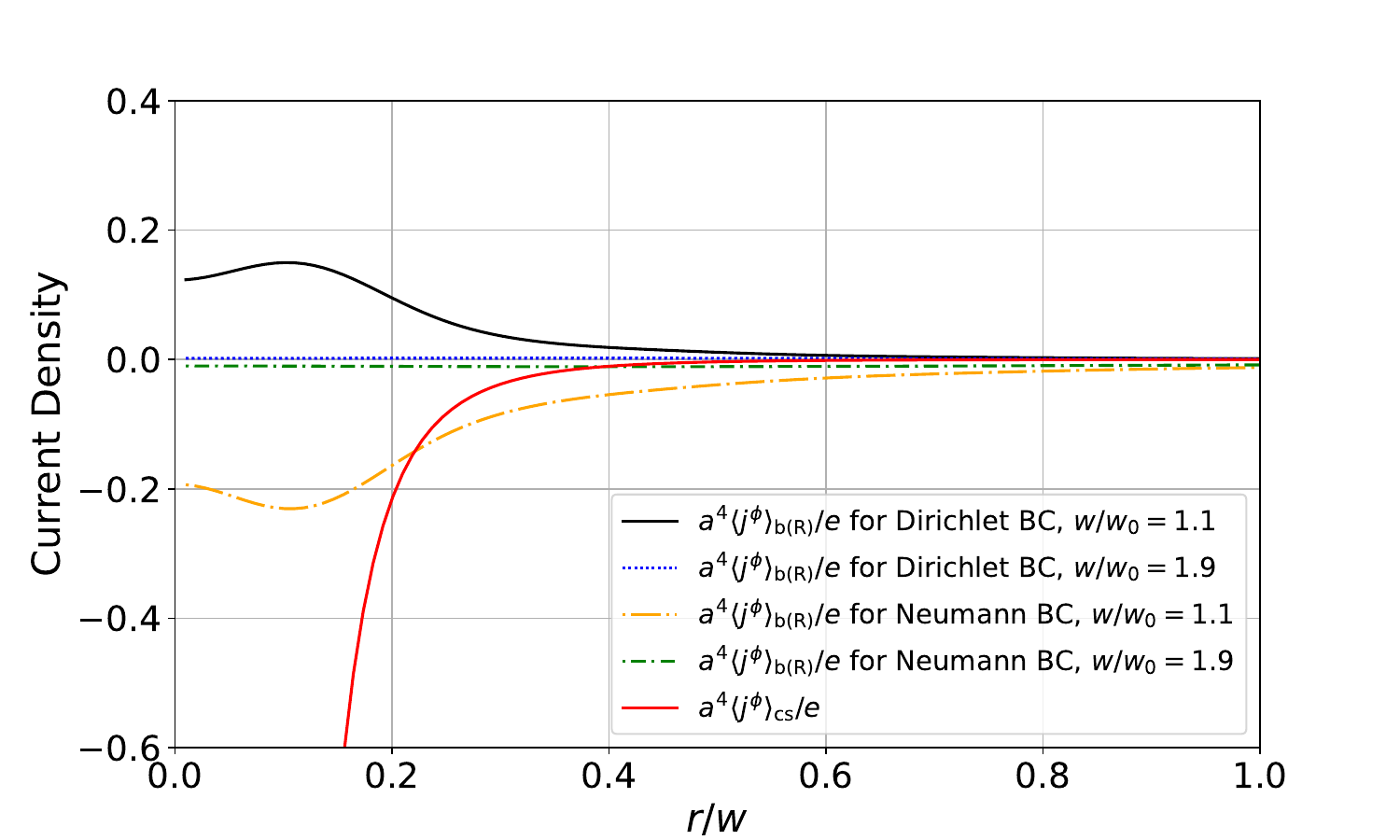}
		\quad
		\includegraphics[scale=0.30]{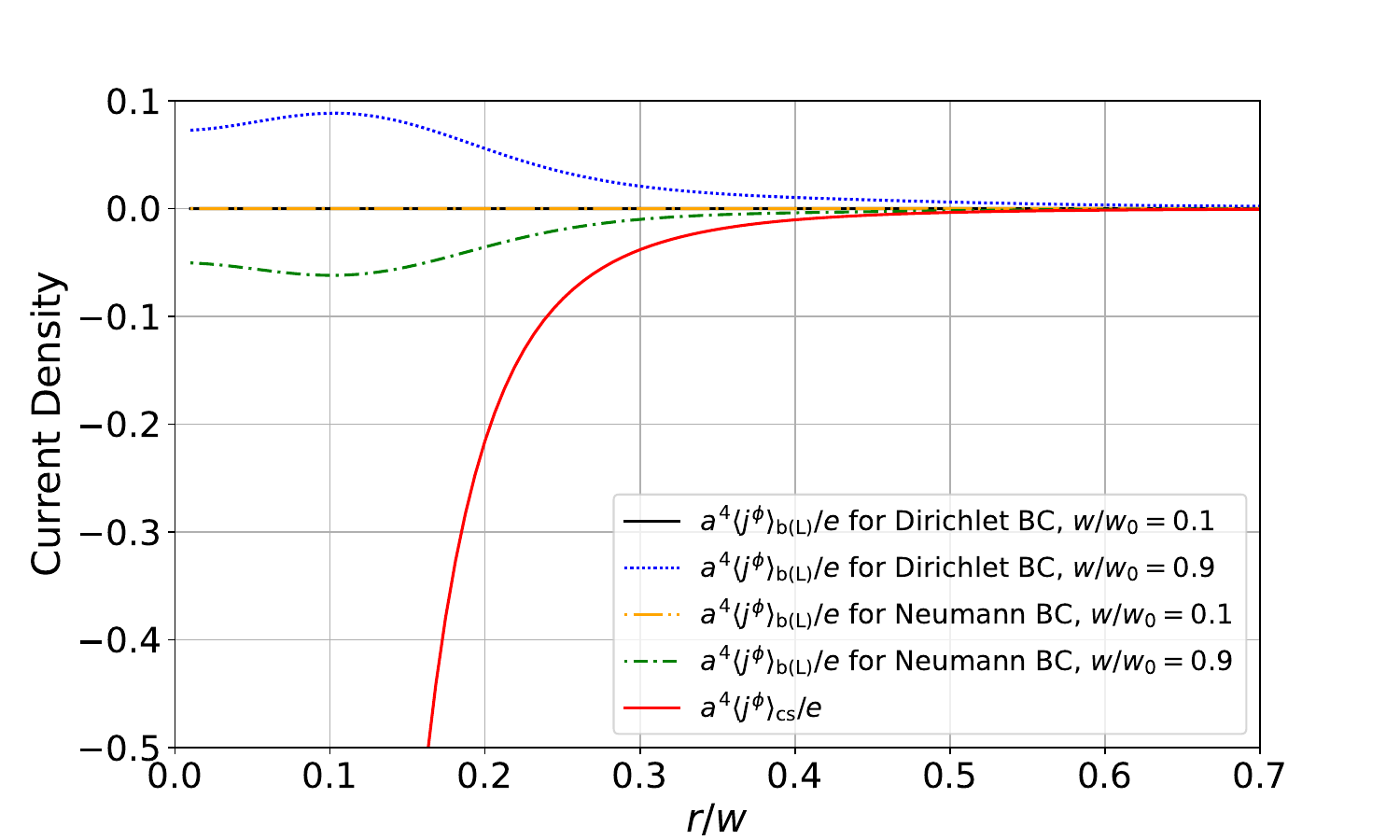}
\caption{The behavior of the azimuthal current density without brane and the brane-induced ones are presented considering $D=3$, $\nu=3/2$, $\alpha_0=0.4$ and $q=1.5$. The plot on the top corresponds to the $R-$region, and the plot on the bottom to the $L-$region. }
		\label{fig1}
	\end{center}
\end{figure}

	\section{Concluding remarks}
	We have studied the combined effects of curvature, conical topology and presence of a brane on the  brane  induced current associated with a massive charged scalar field propagating in a high-dimensional AdS spacetime in the presence of an idealized carrying-magnetic-flux cosmic string. The  latter considered perpendicular to the brane which is parallel to the AdS boundary. On the brane we impose that the field operator obeys  Robin boundary conditions.   The brane divides the manifold in two regions, $R$ and $L$. The boundary induced azimuthal  currents in both regions were exactly obtained in (\ref{azimu_brane_R}), for $R$ region, and (\ref{azimu_brane_L}), for $L$ region. These being odd functions of $\alpha_0$, the fractional part of the ratio $\Phi/\Phi_0$, that is an Aharonov-Bohm-like effect; moreover, their  magnitude depend on the distance with respect to the brane. For  $w/w_0\gg1$, $\langle j^{\phi}\rangle_{b(R)}$ decays as $(w_0/w)^{2\nu}$, (\ref{azimu_brane_R_1}), while for $w\ll w_0$,  $\langle j^{\phi}\rangle_{b(L)}$ decays as $(w/w_0)^{2\nu}$, (\ref{azimu_brane_L-asymp}). This is a Casimir-like effect.  Also we Section \ref{total} we present two plots comparing the behaviors of the current induced by the cosmic string only, with the ones induced by the brane, for $R-$ and $L-$ regions. By these plots we can see that, near the string, the current in the absence of brane is dominant.

	%\begin{thebibliography}{000} %for 3 digits
	%\begin{thebibliography}{00}  %for 2 digits
	
\end{document}